\documentclass[aps,prev,twocolumn,preprintnumbers,floatfix,nofootinbib]{revtex4-1}
\pdfoutput=1
\usepackage{graphicx}
\usepackage{epstopdf}
\usepackage{mathrsfs}
\usepackage{amssymb}
\usepackage{verbatim}
\usepackage{color}
\usepackage{multirow}

\usepackage{amsmath}

\newcommand{\beq}{\begin{equation}}
\newcommand{\eeq}{\end{equation}}
\newcommand{\ga}{\lower.7ex\hbox{$\;\stackrel{\textstyle>}{\sim}\;$}}
\newcommand{\la}{\lower.7ex\hbox{$\;\stackrel{\textstyle<}{\sim}\;$}}

\usepackage{hyperref}
\usepackage{amsmath,bm}
\usepackage{physics}

\hypersetup{
    colorlinks = true,
    citecolor = {blue},
    linkcolor = {blue},
    urlcolor = {blue},
}

\begin{document}

\def\jcap{\ref@jnl{J. Cosmology Astropart. Phys.}}

\vspace{0.5cm}
\title{BICEP/{\it Keck} Constraints on Attractor Models of Inflation and Reheating}

\author{John~Ellis$^{a}$}
\author{{Marcos A.~G.~Garcia}$^{b}$}
\author{Dimitri~V.~Nanopoulos$^{c}$}
\author{Keith~A.~Olive$^{d}$}
\author{Sarunas~Verner$^{d}$}

\affiliation{
\vspace{5mm}
$^a$Theoretical Particle Physics and Cosmology Group, Department of
  Physics, King's~College~London, London WC2R 2LS, United Kingdom;\\
Theoretical Physics Department, CERN, CH-1211 Geneva 23, Switzerland;\\
NICPB, R\"avala pst. 10, 10143 Tallinn, Estonia}
\affiliation{$^b$Departamento de F\'isica Te\'orica, Instituto de F\'isica, Universidad Nacional Aut\'onoma de M\'exico, A.P.~20-364, Ciudad de M\'exico 01000, Mexico}
\affiliation{$^c$George P. and Cynthia W. Mitchell Institute for Fundamental Physics and Astronomy, Texas A\&M University, College Station, TX 77843, USA;\\
Astroparticle Physics Group, Houston Advanced Research Center (HARC),
Mitchell Campus, Woodlands, TX 77381, USA;\\
Academy of Athens, Division of Natural Sciences, Athens 10679, Greece}
 \affiliation{$^d$William I. Fine Theoretical Physics Institute, School of
 Physics and Astronomy, University of Minnesota, Minneapolis, MN 55455,
 USA}

\begin{abstract}
Recent BICEP/{\it Keck} data on the cosmic microwave background, 
in combination with previous
WMAP and {\it Planck} data, impose strong new constraints on the tilt
in the scalar perturbation spectrum, $n_s$, 
as well as the tensor-to-scalar ratio, $r$. 
These constrain the number of $e$-folds of
inflation, $N_*$, the magnitude of the inflaton coupling to matter,
$y$, and the reheating temperature, $T_{\rm reh}$, which we evaluate
in attractor models of inflation as formulated in
no-scale supergravity. The 68\% C.L. region of $(n_s, r)$
favours large values of $N_*, y$, and $T_{\rm reh}$ that
are constrained by the production of gravitinos and supersymmetric
dark matter.

\begin{center}
{\tt KCL-PH-TH/2021-90, CERN-TH-2021-199, ACT-06-21, MI-HET-769, UMN-TH-4107/21, FTPI-MINN-21/26} 
\end{center}

\end{abstract}

\maketitle

\section{Introduction}

Successive releases of data on perturbations in the cosmic microwave
background (CMB)~\cite{Planck2018} have provided increasingly strong upper limits on the
tensor-to-scalar ratio, $r$, and hence sharpened focus on models of
inflation that favour small values of $r$, such as the original
Starobinsky model~\cite{Staro} that predicts $r \sim 0.004$ for 55 $e$-folds. 
The recent release of the BICEP/{\it Keck}~\cite{BICEP2021} data has followed this
trend, imposing the bound $r_{0.05} < 0.036$ at the 95\% C.L. where the subscript denotes the pivot scale in Mpc${^{-1}}$. Moreover,
the combination of WMAP, {\it Planck} and BICEP/{\it Keck} data
constrains the scalar tilt to the limited range 
$0.958 < n_s < 0.975$ at the 95\% C.L. for $r = 0.004$. 
A further analysis by \cite{Tristram:2021tvh} used BB autocorrelation data from \cite{PR4} and allowed a free reionization optical depth,
and obtained a lower limit on the scalar-to-tensor ratio to $r_{0.05} < 0.032$, with a slightly relaxed range on the spectral tilt 
$0.956 < n_s < 0.974$ at the 95\% C.L. for $r = 0.004$.

The Starobinsky model is not alone in accommodating the upper limit on $r$. For example, Higgs inflation predicts a similar value of $r$~\cite{higgsinf}, and similar potentials appear naturally in the context of supergravity, including no-scale supergravity \cite{no-scale1, no-scale2}. In particular, the simplest no-scale
supergravity models characterized by a K{\" a}hler potential of the form
$K \; = \; -3 \, \ln(T + \overline{T} - |\phi|^2/3)$, where $T$ and $\phi$ are complex scalar fields, predict a Starobinsky-like value of $r$~\cite{ENO6},
but the no-scale supergravity framework can also accommodate other possibilities \cite{building}.

For example, generalizing $- 3 \to -3 \alpha$ as the coefficient of the logarithm
modifies the prediction for $r$ by a factor $\alpha$, as was first pointed out in~\cite{ENO7}
and subsequently in~\cite{KLR}. Such a modification of the simplest no-scale model is a natural
possibility in compactified string models, where $T$ may be interpreted as the
volume modulus~\cite{Witten}, which is a product of three independent compactification moduli 
$T_i: i = 1, 2, 3$. Models in which inflation is driven by one (two) of these moduli
correspond to $\alpha = 1/3 \ (2/3)$~\cite{ENO7}. Larger values of $\alpha$ are
also possible, since string compactifications also have complex structure moduli that
can contribute to the inflationary dynamics~\cite{Kallosh:2013hoa}.

A common feature of these no-scale supergravity models is a quadratic singularity in the kinetic
term for the inflaton. This feature leads generically to an effective potential for the
canonically normalized inflaton field with a plateau that leads to a
quasi-de Sitter inflationary epoch similar to that in Starobinsky inflation. This
property was abstracted from the no-scale models in~\cite{Kallosh:2013hoa}, where they were baptized
``attractor'' models. Two specific types of attractor potential can be 
distinguished~\cite{ENO7,KLR,T-model,ENOV3}:~\footnote{We note that $\alpha$-Starobinsky models are also known as $E$ models \cite{e-m}.}
\begin{align}
\label{eq:alphastaro}
V & \; = \; \frac{3}{4} \lambda  M_{P}^{4}\left(1-e^{-\sqrt{\frac{2}{3 \alpha}} \frac{\varphi}{M_{P}}}\right)^{2},  & \nonumber \\
& \qquad \qquad \qquad \qquad \; \; (\text{$\alpha$-Starobinsky~\cite{ENO7,KLR,ENOV3}})&  
\end{align}
\begin{align}
\label{eq:tmodel}
V & \; = \; \frac{3}{4}  \lambda M_P^4 \tanh ^{2} \left(\frac{\varphi}{\sqrt{6 \alpha}M_P } \right), & \nonumber \\
& \qquad \qquad \qquad \qquad \qquad \qquad (\text{T Model~\cite{T-model}}) &
\end{align}
where $\varphi$ is the canonically normalized inflaton field, $M_P = \frac{1}{\sqrt{8 \pi G}} \simeq 2.435 \times 10^{18} \, \rm{GeV}$ is the reduced Planck mass, and $\lambda$ is the potential scale determined from the CMB normalization and the inflaton field value at horizon crossing.~\footnote{The normalization of the potentials is chosen so that the inflaton normalization scale coincides in both cases, and is given by Eq.~(\ref{eq:infnorm}). This choice does not affect the CMB observables $n_s$ and $r$.} 
For the attractor models discussed here, increasing the value of $\alpha$ reduces the flatness of the plateau at the inflaton field value at the horizon crossing of the CMB scale, $\varphi_*$, which affects the cosmological observables $n_s$ and $r$.
It was argued in~\cite{ENO7,KLR,T-model,rs,ENOV3,building} that broad classes of 
attractor models lead to identical predictions of $n_s$ and $r$ in the limit of a large number of $e$-folds, $N_*$.\footnote{We note that
the potentials (\ref{eq:alphastaro}) and (\ref{eq:tmodel}) are identical at zeroth and first order in $e^{-\sqrt{\frac{2}{3 \alpha}} \frac{\varphi}{M_{P}}}$, but differ
at higher orders and so make different predictions when $\varphi/\sqrt{\alpha} = {\cal O}(M_P)$. One could in principle consider other attractor potentials that are
also equivalent at zeroth and first order, but these are the options commonly considered in the literature.}
In the context of supergravity,
the parameter $\alpha$ determines the curvature of the internal K\"ahler manifold: $R = 2/\alpha$.\footnote{In general, the K\"ahler curvature $R$ depends on the total number, $n$, of chiral fields describing the theory~\cite{no-scale1, no-scale2, EKN1,ENOV3}, $R = n(n+1)/3\alpha$, and this result holds for two chiral fields, which is the minimal number needed to construct a plateau-like potential in no-scale supergravity~\cite{ENO7}.}  

In this paper we explore the impact of the latest BICEP/{\it Keck}/WMAP/{\it Planck} constraints in the $(n_s, r)$ plane on the $\alpha$-Starobinsky and T model inflationary attractors (see also~\cite{KL2021}) from both \cite{BICEP2021} and \cite{Tristram:2021tvh}. From the analysis in \cite{BICEP2021}, we find that the region of CMB parameters favoured at the 68\% C.L. by the combination of CMB data favours $N_{0.05} \gtrsim 50.9 \, (52.6)$ in the $\alpha$-Starobinsky (T models), corresponding to an inflaton decay coupling 
$y \gtrsim 1.7 \times 10^{-6} (1.7 \times 10^{-4})$ for $\alpha = 1$, with an order of magnitude sensitivity to $\alpha \in (0.1, 5)$.~\footnote{The corresponding 95\% limits are $N \gtrsim 45.9 (47.5)$ and $y \gtrsim 3.8 \times 10^{-13} \, (3.6 \times 10^{-11})$, respectively.} 
In contrast, the analysis in \cite{Tristram:2021tvh} yields substantially weaker bounds, $N_{0.05} \gtrsim 47.9 \, (49.4)$ in the $\alpha$-Starobinsky (T models), corresponding to an inflaton decay coupling 
$y \gtrsim 1.9 \times 10^{-10} \, (1.2 \times 10^{-8})$ for $\alpha = 1$~\footnote{In this case, the corresponding 95\% limits are $N \gtrsim 42.9 (44.6)$ and $y \gtrsim 2.8 \times 10^{-17} (4.0 \times 10^{-15})$, respectively.}. 
Additionally, supergravity models must avoid overproducing gravitinos and supersymmetric dark matter~\cite{ego,ENOV4}. We find that based on \cite{BICEP2021}, $\alpha$-Starobinsky models that respect these constraints fall inside the region favoured by the CMB data at the 68\% C.L. only for $\alpha \in (0.67, 12)$, and that T models fall inside this region only for $\alpha \in (1.3, 5.1)$. At the 95\% C.L. these range are (0, 26) and (0, 11), respectively. Based on \cite{Tristram:2021tvh}, the 68\% C.L. range is (0.4, 12) and (0.5, 7) for the $\alpha$-Starobinsky and T models, respectively, and the 95\% C.L. ranges are (0, 24) and (0, 12).~\footnote{Here the lower bound $\alpha > 0$ arises because $\alpha = 0$ leads to a completely flat potential that is not suitable for inflation.}

\section{Inflationary Dynamics}

The dynamics of the inflaton is characterized by the action
\begin{equation}    
    \label{eq:action}
    \mathcal{S} \; = \; \int d^4 x  \sqrt{-g} \left[\frac{M_P^2}{2} R + \frac{1}{2} \partial_{\mu} \varphi \partial^{\mu} \varphi - V(\varphi) \right] \, ,
\end{equation}
where the effective scalar potential is given by Eq. (\ref{eq:alphastaro}) or (\ref{eq:tmodel}).
We use for our analysis the conventional slow-roll parameters,
which are given in single-field inflationary models by
\begin{equation}\label{eq:epseta}
\epsilon \equiv \frac{1}{2} M_{P}^{2}\left(\frac{V^{\prime}}{V}\right)^{2} \, , \qquad \eta \equiv M_{P}^{2}\left(\frac{V^{\prime \prime}}{V}\right) \, ,
\end{equation}
where the prime denotes a derivative with respect to the inflaton field, $\varphi$.  In the slow-roll approximation, the number of $e$-folds can be computed using
\begin{equation}
\label{eq:efolds}
N_{*} \simeq \frac{1}{M_{P}^{2}} \int_{\varphi_{\mathrm{end}}}^{\varphi_{*}} \frac{V(\varphi)}{V^{\prime}(\varphi)} d \varphi \simeq \int_{\varphi_{\mathrm{end}}}^{\varphi_{*}} \frac{1}{\sqrt{2 \epsilon}} \frac{d \varphi}{M_{P}} \, ,
\end{equation}
where $k_* \; = \; 0.05 \, \rm{Mpc}^{-1}$ is the pivot scale used in the {\it Planck} analysis. The end of inflation occurs when $\ddot{a} = 0$, i.e., $\dot{\varphi}_{\rm{end}}^2 = V(\varphi_{\rm{end}})$.

The principal CMB observables, namely, the scalar tilt, $n_s$, the tensor-to-scalar ratio, $r$, and the amplitude of the curvature power spectrum, $A_S$, can be expressed as follows
in terms of the slow-roll parameters: 
\begin{align}
    \label{eq:spectrtilt}
    n_{s} \; &\simeq \; 1-6 \epsilon_{*}+2 \eta_{*} \, , \\
    \label{eq:sclrtotens}
    r \; &\simeq \; 16 \epsilon_{*} \, , \\
    \label{eq:powerspectr}
    A_{S *} \; &\simeq \; \frac{V_{*}}{24 \pi^{2} \epsilon_{*} M_{P}^{4}} \, , 
\end{align}
where $V_* = V(\varphi_*)$ and $A_{S *} \simeq 2.1 \times 10^{-9}$~\cite{Planck2018}. In the large $N_*$ limit, the inflationary attractor potentials~(\ref{eq:alphastaro}) and~(\ref{eq:tmodel}) predict~\cite{ENO7}
\begin{equation}
\label{cmbpredictions}
n_{s}  \; \simeq \; 1-\frac{2}{N_{*}}, \qquad r \; \simeq \; \frac{12 \alpha}{N_{*}^{2}} \, , 
\end{equation}
where the approximation holds for $\alpha \lesssim \mathcal{O}(1)$
in $\alpha$-Starobinsky models, and the full analytical expression can be found in~\cite{ENOV4}.

Using expression~(\ref{eq:efolds}), we can calculate the approximate value of the inflaton field at the horizon exit scale $k_*$~\cite{EGNO5} when $\alpha = 1$,
\begin{align}
\label{alphastarophistar} \nonumber
\frac{\varphi_{*}}{M_P} & \; \simeq \; \sqrt{\frac{3}{2}} \left[1 + \frac{3}{4N_*-3} \right]\\
&\qquad \times \ln\left(\frac{4N_*}{3} + e^{\sqrt{\frac{2}{3}} \frac{\varphi_{\rm{end}}}{M_P}} - \sqrt{\frac{2}{3}} \frac{\varphi_{\rm{end}}}{M_P}\right) , & \nonumber \\
& \qquad \qquad \qquad \qquad \qquad \qquad (\text{$\alpha$-Starobinsky}) & \\
\label{tmodelphistar}
\frac{\varphi_{*}}{M_P} & \; \simeq \;  \sqrt{\frac{3}{2}}  \cosh^{-1} \left(\frac{4N_*}{3} + \cosh\left(\sqrt{\frac{2}{3}} \frac{\varphi_{\rm{end}}}{M_P} \right) \right)\, , & \nonumber \\
& \qquad \qquad \qquad \qquad \qquad \qquad \qquad (\text{T Model}) &
\end{align}
with
\begin{align}
\label{alphastarophiend}
\frac{\varphi_{\rm{end}}}{M_P} & \; \simeq \; \sqrt{\frac{3}{2}} \ln\left[\frac{2}{11} \left(4 + 3\sqrt{3} \right)\right] , & \nonumber \\
& \qquad \qquad \qquad \qquad \qquad \quad (\text{$\alpha$-Starobinsky}) & \\
\label{tmodelphiend}
\frac{\varphi_{\rm{end}}}{M_P} & \; \simeq \;  \sqrt{\frac{3}{2}}  \ln\left[\frac{1}{11} \left(14 + 5\sqrt{3} \right)\right] ,
\; \; (\text{T Model}) &
\end{align}
where $\varphi_{\rm{end}}$ was calculated using the expression~$\epsilon = (1 + \sqrt{1 -\eta/2})^2$, and the full analytical approximations for $\varphi_*$ and $\varphi_{\rm{end}}$ can be found in Appendix~\ref{appA}, where they are given by Eqs.~(\ref{eq:phiendE})-(\ref{eq:phistarT}).  
Combining the expressions above with  expression~(\ref{eq:powerspectr}) for the curvature power spectrum, we find that the inflaton normalization scale is proportional to $\lambda$, which is in turn proportional to $\alpha$ and given by
\begin{equation}
\label{eq:infnorm}
\lambda  \; \simeq \; \frac{24 \alpha \pi^2  A_{S *} }{N_*^2} \, .
\end{equation}
We now calculate the number of $e$-folds, $N_*$, assuming that there is no additional entropy injection between the end of reheating and when the horizon scale $k_*$ reenters the horizon \cite{Martin:2010kz, LiddleLeach}:
\begin{equation}
\begin{aligned}
\label{eq:nstarreh}
N_{*} \; &= \; \ln \left[\frac{1}{\sqrt{3}}\left(\frac{\pi^{2}}{30}\right)^{1 / 4}\left(\frac{43}{11}\right)^{1 / 3} \frac{T_{0}}{H_{0}}\right]-\ln \left(\frac{k_{*}}{a_{0} H_{0}}\right) \\
& -\frac{1}{12} \ln g_{\mathrm{reh}} \\
&+\frac{1}{4} \ln \left(\frac{V_{*}^{2}}{M_{P}^{4} \rho_{\mathrm{end}}}\right) +\frac{1-3 w_{\mathrm{int}}}{12\left(1+w_{\mathrm{int}}\right)} \ln \left(\frac{\rho_{\mathrm{rad}}}{\rho_{\text {end }}}\right) 
\, ,
\end{aligned}
\end{equation}
where the present Hubble parameter and photon temperature are given by $H_0 = 67.36 \, \rm{km \, s^{-1} \, Mpc^{-1}}$~\cite{Planck:2018vyg} and $T_0 = 2.7255 \, \rm{K}$~\cite{Fixsen:2009ug}. Here, $\rho_{\rm{end}}$ and $\rho_{\rm{rad}}$ are the energy density at the end of inflation and at the beginning of the radiation domination era when $w = p/\rho = 1/3$, respectively, $a_0 = 1$ is the present day scale factor, $g_{\rm{reh}} =  915/4$ is the effective number of relativistic degrees of freedom in the minimal supersymmetric standard model (MSSM) at the time of reheating, and the equation of state parameter averaged over the $e$-folds during reheating is
\begin{equation}
w_{\mathrm{int}} \equiv \frac{1}{N_{\mathrm{rad}}-N_{\mathrm{end}}} \int_{N_{\mathrm{end}}}^{N_{\mathrm{rad}}} w(n) \, d n \, .
\end{equation}
Using the numerical values given above with the Planck pivot scale $k_* \; = \; 0.05 \, \rm{Mpc}^{-1}$,~\footnote{We note that when we calculate the tensor-to-scalar ratio $r_{\rm{0.002}}$ numerically, we evaluate $N_*$ at the pivot scale $k_* = 0.002 \, \rm{Mpc}^{-1}$.} we find the
following value for the sum of the first two lines in (\ref{eq:nstarreh}):
$N_*  \simeq  61.04 + \cdots$. 
Mechanisms for producing a baryon asymmetry (such as leptogenesis) are simplified when $T_{\rm reh} \gtrsim$ the electroweak scale. Accordingly, we also display results for a reheating temperature $T_{\rm reh} = T_{\rm EW} \sim 100$ GeV, whilst acknowledging that lower reheating temperatures are possible. For $T_{\rm reh} = T_{\rm EW}$ we take the Standard Model value for $g_{\rm reh} = 427/4$, and find $N_{\rm EW} = 61.10 + \cdots$.
The minimum reheating temperature that is compatible with Big Bang Nucleosynthesis (BBN) is $T_{\rm{reh}} \gtrsim {\cal O}(1)\; \rm{MeV}$. Using $T_{\rm{BBN}} = 2 \; \rm{MeV}$ in
our numerical analysis, corresponding to $g_{\rm{reh}} = 10.75$, the sum of the first two lines of (\ref{eq:nstarreh})
takes the following numerical value:
$N_{\rm{BBN}} \simeq 61.29 + \cdots.$ 

To calculate the values of $N_*$, $N_{\rm EW}$ and $N_{\rm{BBN}}$ numerically, we use the following equations that govern the cosmic background dynamics:
\begin{align}
\label{eq:dyn1}
\dot{\rho}_{\varphi}+3 H \rho_{\varphi}& =-\Gamma_{\varphi} \rho_{\varphi} \, , \\
\label{eq:dyn2}
\dot{\rho}_{r}+4 H \rho_{r} &=\Gamma_{\varphi} \rho_{\varphi} \, , \\
\label{eq:dyn3}
\rho_{\varphi}+\rho_{r} &=3 M_{P}^{2} H^{2} \, ,\\
\frac{d}{dt}{(N w_{\rm int})} &= H w \,, 
\label{eq:dyn4}
\end{align}
where $\rho_{\varphi}$ and $\rho_{r}$ are the energy densities of the inflaton and produced radiation, respectively, and $\Gamma_{\varphi}$ is the inflaton decay rate given by
\begin{equation}
    \label{eq:decayrate}
    \Gamma_{\varphi} \; = \; \frac{y^2}{8 \pi} m_{\varphi} \, ,
\end{equation}
where $y$ is a Yukawa-like coupling, and we find the following masses
in the inflationary attractor potentials~(\ref{eq:alphastaro}) and~(\ref{eq:tmodel}): 
\begin{align}
\label{eq:infmass1}
m_{\varphi} & \; = \; \sqrt{\frac{\lambda}{\alpha}} M_P \, ,  &(\text{$\alpha$-Starobinsky})  \\
\label{eq:infmass2}
m_{\varphi} & \; = \; \frac{1}{2} \sqrt{\frac{\lambda}{\alpha}} M_P \, . &(\text{T Model}) 
\end{align}

\section{Reheating}

The reheating process occurs after the end of inflation in a matter-dominated background. As the inflaton starts to decay, the dilute plasma reaches a maximum temperature, $T_{\rm{max}}$~\cite{Giudice:2000ex, Ellis:2015jpg}, and subsequently starts falling as $T \propto a^{-3/8}$. The reheating temperature is defined through~\cite{GKMO,Pallis:2005bb}
\begin{equation}
    \label{eq:reheating}
    \frac{\pi^2 g_{\rm{reh}} T_{\rm{reh}}^4}{30} \; = \; \frac{12}{25} \left(\Gamma_{\varphi} M_P \right)^2 \, ,
\end{equation}
when the energy density of the inflaton is equal to the energy density of radiation,
corresponding to
\begin{equation}
    T_{\rm{reh}} \; \simeq \; 1.9 \times 10^{15} \, {\rm{GeV}} \cdot y \cdot g_{\rm{reh}}^{-1/4}  \left(\frac{m_{\varphi}}{3 \times 10^{13} \, \rm{GeV}} \right)^{1/2} \, .
\end{equation}
In order to evaluate the constraint on $T_{\rm reh}$
from overproduction of supersymmetric dark matter in scenarios where the gravitino is lighter than $T_{\rm reh}$, we use the expression~\cite{Ellis:2015jpg, Eberl:2020fml}\footnote{We use here
an analytical approximation since there is only a 0.03 \% difference between the analytical and fully numerical calculation.}
\begin{equation}
\label{gravitinoprodn}
Y_{3 / 2}(T) \; = \; 0.00336\left(1 + 0.51 \frac{m_{1 / 2}^{2}}{m_{3 / 2}^{2}}\right)\left(\frac{\Gamma_{\varphi}}{M_{p}}\right)^{1 / 2} \, ,
\end{equation}
where $Y_{3/2} \equiv n_{3/2}/n_{\rm{rad}}$ is the gravitino yield, $n_{\rm rad} = \zeta(3) T^3/\pi^2$, $m_{3/2}$ the gravitino mass, and $m_{1/2}$ the gluino mass \cite{ekn,enor,bbb}.
Disregarding the term $m_{1/2}^2/m_{3/2}^2$
in (\ref{gravitinoprodn}) and using the observed dark matter density today, $\Omega_{\rm CDM} h^2 \simeq 0.12$, we find the following upper limit on the Yukawa-like inflaton coupling, assuming that the gravitino decays after the lightest supersymmetric particle (LSP) decouples,
\begin{equation}
    |y| < 9.2  \times 10^{-8} \sqrt{\frac{M_P}{m_{\varphi}}} \left( \frac{100  \, \rm{GeV}}{m_{\rm{LSP}}}\right) \, ,
\end{equation}
where $m_{\rm{LSP}}$ is the mass of the LSP and the inflaton masses for the different inflationary attractor potentials are given by Eqs.~(\ref{eq:infmass1}) and (\ref{eq:infmass2}).\footnote{If the gravitino is the LSP, the second term in the
brackets in (\ref{gravitinoprodn}) must be taken into account, and the constraint
on $y$ depends on the ratio $m_{1/2}/m_{3/2}$.} We note that, since $m_{\varphi} \propto 1/\sqrt{\alpha}$, $|y| \propto \alpha^{1/4}$.\footnote{For another recent analysis of
gravitino constraints in light of the BICEP/{\it Keck} results, see~\cite{KO2021}.}

In high-scale supersymmetry models in which the gravitino mass may be significantly larger than the electroweak scale
and the other supersymmetric particles are heavier than the inflaton, the gravitino, which is now the LSP, is pair-produced via its
longitudinal components \cite{Dudas:2017rpa}. In such a scenario, we find~\cite{Garcia:2018wtq}
\begin{equation}
\begin{aligned}
\Omega_{3 / 2} h^{2} \; \simeq \; & 0.12\left(\frac{|y|}{3.0 \times 10^{-7}}\right)^{19 / 5}\left(\frac{m_{\varphi}}{3 \times 10^{13} \, \mathrm{GeV}}\right)^{67 / 10} \\
& \times\left(\frac{0.1 \, \mathrm{EeV}}{m_{3 / 2}}\right)^{3}\left(\frac{0.030}{\alpha_{3}}\right)^{16 / 5} \, ,
\end{aligned}
\end{equation}
where $m_{3/2}$ is the gravitino mass and $\alpha_3$ is the strong coupling. Using the observed dark matter abundance today to constrain $\Omega_{3/2} h^2$, we find that avoiding overproduction of dark matter imposes the following bound:
\begin{equation}
    |y| < 6.6 \times 10^{-16} \, \left(\frac{M_P}{m_{\varphi}} \right)^{67/38} \left(\frac{m_{3/2}}{0.1 \, \rm{EeV}} \right)^{15/19} \, .
\end{equation}
We note that in a non-supersymmetric theory there would, in general, be a lower limit on $y$ due to the fact that it generates radiative corrections $\propto y^4$ in
the effective inflaton potential~\cite{DreesXu}. However, this is not the case in supersymmetric models such as those discussed above, where these radiative corrections cancel down to the level
of the relatively small supersymmetry-breaking effects~\cite{ENOT}.

\section{Results}

We solve the cosmic background equations~(\ref{eq:dyn1})-(\ref{eq:dyn4}) numerically to
determine the number of $e$-folds $N_*$, $N_{\rm EW}$, and $N_{\rm{BBN}}$. In the $\alpha = 1$ case, the procedure of calculating the analytical approximations for $N_*$ is given in Appendix~\ref{appA} (see Eqs.~(\ref{eq:nEmodapprox}) and~(\ref{eq:nTmodapprox})). The full numerical computation of the CMB observables is discussed in Appendix~\ref{appB}.

Figure~\ref{fig:results} summarizes our numerical results based on the analysis of \cite{BICEP2021}:
those for $\alpha$-Starobinsky models are shown in
the upper pair of panels and those for T models in
the lower pair. For each of the two models, we derive limits on $N_*$ from the requirements that $T_{\rm reh} > 2$ MeV (100 GeV) and the supersymmetric relic density when $m_{\rm LSP} = 100$ GeV. The former gives a lower limit to $N_*$, while the latter gives an upper limit. We also derive the corresponding limits on $y$.
These are compared to the 68\% and 95\% C.L. limits on $N$ and $y$ from the BICEP/{\it Keck} constraints on $n_s$. For $\alpha = 1$, we find the following limits: 
\begin{align}
\label{eq:limits1}
&\qquad \qquad \text{$\alpha$-Starobinsky}: \nonumber \\
&41.8 (45.6) < \, N_* \, < 51.8,& \nonumber \\
&1.7 \times 10^{-18} (1.6 \times 10^{-13}) < \, |y| \, < 2.6 \times 10^{-5},& \nonumber \\
&N_{68 \%}  = 50.9, \quad N_{95 \%} = 45.9,& \nonumber \\
&T_{{\rm{reh}, \,} 68 \%} = 8.7 \times 10^{8} \, {\rm{GeV}}, \quad T_{{\rm{reh}, \,} 95 \%} = 2.4 \times 10^{2} \, \rm{GeV},& \nonumber \\
&y_{68 \%} = 1.7 \times 10^{-6}, \quad y_{95 \%} = 3.8 \times 10^{-13},& \\
&~~& \nonumber \\
\label{eq:limits2}
& \qquad \qquad \text{T Model}: \nonumber \\
&42.0 (45.8) < \, N_* \, < 52.1,& \nonumber \\
&2.3 \times 10^{-18} (2.2 \times 10^{-13}) < \, |y| \, < 3.6 \times 10^{-5},& \nonumber \\
&N_{68 \%} = 52.6, \quad N_{95 \%} = 47.5,& \nonumber \\
&T_{{\rm{reh}, \,} 68 \%} = 5.9 \times 10^{10} \, {\rm{GeV}}, \quad T_{{\rm{reh}, \,} 95 \%} = 1.4 \times 10^{4} \, \rm{GeV},& \nonumber \\
&y_{68 \%} = 1.7 \times 10^{-4}, \quad y_{95 \%} = 3.6 \times 10^{-11}.& 
\end{align}
We note that the first two lines do not depend on the BICEP/{\it Keck} constraints, since these limits are derived from the conditions $T_{\rm reh} > 2$ MeV (100 GeV) (smaller limit) and $m_{\rm LSP} = 100$ GeV (larger limit). The dark (light) blue regions in the left panels are the 68 (95) \% C.L. regions of the $(n_s, r_{0.002})$ planes favoured by a global analysis of the CMB and BAO data. 

We also show in the left panels of 
Fig.~\ref{fig:results} dotted contours corresponding to 60
and 50 $e$-folds, solid lines corresponding to the maximum number of $e$-folds consistent with $y \le 1$, and the minimum number of $e$-folds
consistent with $T_{\rm reh} > T_{\rm BBN}$ and $T_{\rm EW}$, as well as the dark matter
density constraints for a LSP mass of 100~GeV. The corresponding limit for 
a gravitino mass of $10^8$~GeV in the high-scale supersymmetry case would lie roughly midway between the $m_{\rm LSP} = 100$ GeV and $N_* = 50$ lines. For the $\alpha$-Starobinsky (T models) we shade in red (orange) the preferred region respecting the constraint $T_{\rm reh} > T_{\rm EW}$ and the relic
density constraint with $m_{\rm LSP} = 100$ GeV. In the upper left panel we also show 
lines corresponding to $\alpha = 1$ and 12, the
latter being the largest value allowed at the 68\% C.L.
for $m_{\rm LSP} = 100$~GeV, and $\alpha = 26$, the
largest value allowed at the 95\% C.L. for 
$m_{\rm LSP} = 100$~GeV. We see in the lower left panel
that values of $\alpha \in (1.3, 5.1)$  are consistent with the
data at the 68\% C.L. if $m_{\rm LSP} = 100$~GeV, and
values of $\alpha \le 11$ are allowed at the 95\% C.L.

\begin{figure*}[ht!]
  \centering 
  \includegraphics[width = \textwidth]{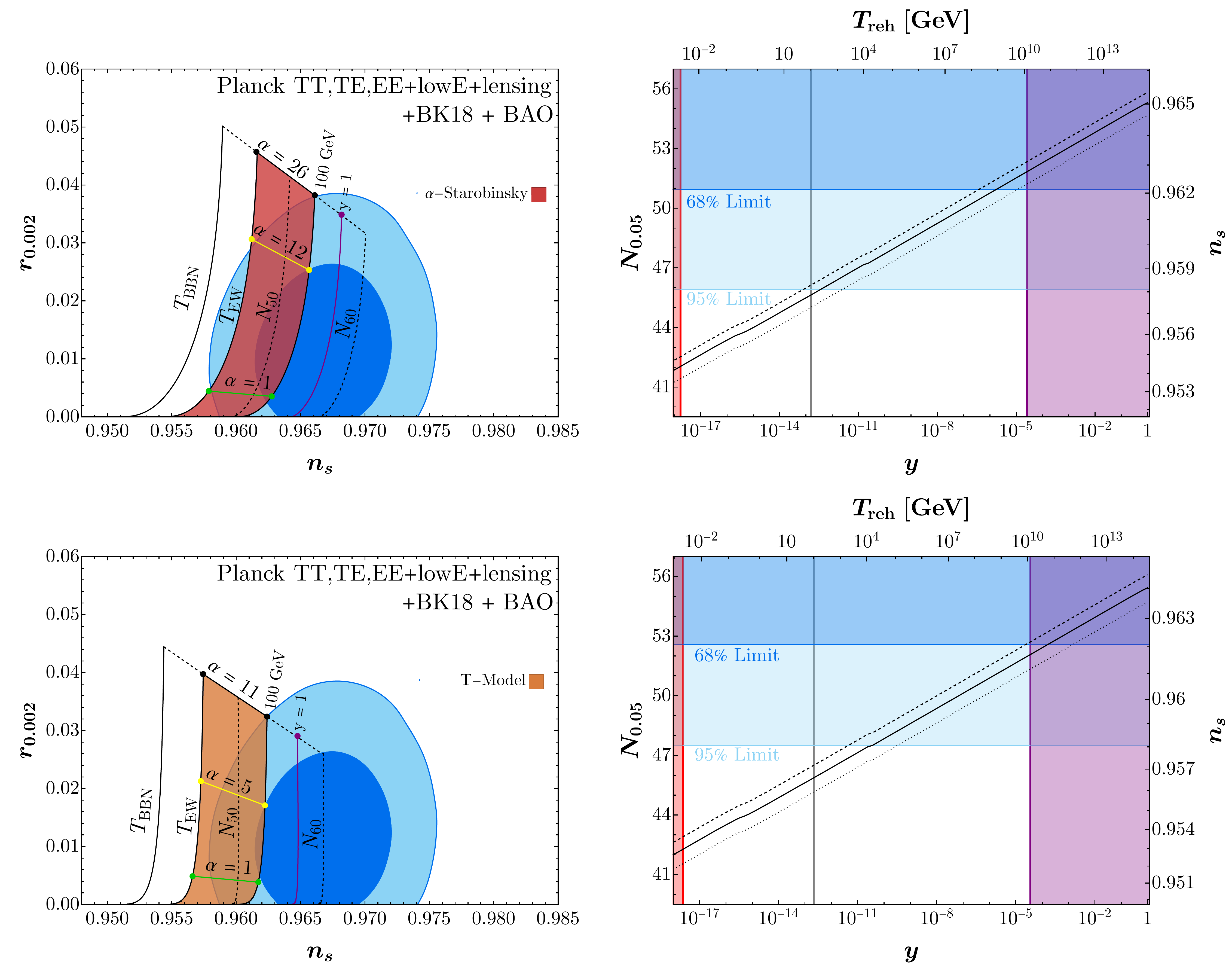}
  \caption{\it Illustrations of the impacts of the BICEP/{\it Keck} and other constraints on $\alpha$-Starobinsky models (upper panels) and T models (lower panels) based on the analysis of \cite{BICEP2021}. 
  The left panels compare the observational 68\% and 95\% C.L. constraints in the $(n_s, r)$ plane (using pivot scales 0.002 for $r$ and 0.05 for $n_s$) with the model predictions for different numbers of $e$-folds
  $N_{50, 60}$, showing also the predictions for an inflaton coupling $y = 1$, the  constraints from $T_{\rm reh} \ge T_{\rm BBN}$ and $T_{\rm EW}$,  and the constraints if the LSP mass is $100 \, \rm{GeV}$.
  The right panels display $(y, N)$ planes (using the pivot scale 0.05), showing the relations between
  $y$ and $T_{\rm reh}$ and between $N$ and $n_s$, and the values $\alpha = 0.1, 1, 5$ (dashed, solid and dotted black lines). We also include lower limits on $y$ from BBN (red line), 
  $T_{\rm reh} = T_{\rm EW}$ (grey line), and gravitino production (purple line) for $\alpha = 1$, which increase for smaller $\alpha$,
  and 68\% and 95\% C.L. lower limits on $N_{0.05}$ from BICEP/{\it Keck} and other data (blue lines). 
  }
  \label{fig:results}
\end{figure*}

The right panels of Fig.~\ref{fig:results} show the
$(y(T_{\rm{reh}}), \, N_{0.05}(n_s))$ planes for the $\alpha$-Starobinsky models and T models. 
The left-most vertical lines (red) correspond to the minimum values of $y$ allowed by BBN, the middle vertical lines (grey)
correspond to $T_{\rm reh} = T_{\rm EW}$, and the right-most vertical lines (purple) correspond to the maximum values allowed
for $m_{\rm{LSP}} = 100$~GeV.
We assume
$\alpha = 1$ when plotting the parameters and constraints.
The constraints would each move to the right (towards larger values of $y$ and $T_{\rm reh}$) with decreasing values of $\alpha$, though their dependences are weak.
The diagonal lines are the
predictions of the $\alpha$-Starobinsky and T models
for $\alpha = 0.1$ (dashed lines), $1$ (solid lines) and 5 (dotted lines).
Finally, we show as horizontal lines the
lower limits on $n_{0.05}$ at the 68 and 95\% C.L.
We see that the 68\% lower
limit of $N_{0.05}$  requires $y > 1.7 \times 10^{-6}$ in the $\alpha$-Starobinsky model and $y > 1.7 \times 10^{-4}$ for the T-Starobinsky model, both for
$\alpha = 1$. This implies a lower limit to the reheating temperature of $8.7 \times 10^{8}$ GeV and $5.9 \times 10^{10}$ GeV for the $\alpha$-Starobinsky models and T models, respectively.  This limit is relaxed at the 95 \% C.L., where the lower limit on the reheating temperature drops to $2.4 \times 10^{2}$ GeV in the $\alpha$-Starobinsky models and $1.4 \times 10^4$ GeV for the T models. 

We assumed in the above analysis that generation of a factor $\Delta$ of entropy subsequent to inflaton decay 
could be neglected. However, this may not be the case, e.g., in models with additional phase transitions
at temperatures between $T_{\rm reh}$ and $T_{\rm EW}$, such as those based on flipped SU(5) GUTs \cite{EGNNO3}.
In this case there would be a modification to the calculation of $N_*$ in Eq.~(\ref{eq:nstarreh}) in the form of an extra term $- \frac13 \ln \Delta$ in the right-hand side. This would in turn modify the left panels of Fig.~\ref{fig:results}, e.g.,
the $T_{\rm BBN}$ and $T_{\rm EW}$ constraints would move to lower $n_s$, as would the $y = 1$ line,
whereas the $N_{50}$ and $N_{60}$ lines would be unchanged, as would the LSP density constraint. 
As entropy generation would allow a higher initial gravitino abundance, and thus a higher reheating temperature,
the contribution to $N_*$ from reheating is exactly compensated by the contribution from $\Delta$. In addition, the
lines of fixed $\alpha$ are unchanged. The net result would be to expand the favoured regions of the $(n_s, r_{0.002})$
planes towards lower values of $n_s$, while keeping the same overlaps with the regions of the planes favoured
by the BICEP/{\it Keck} and other constraints at the 68\% C.L. However, this would require higher reheating 
temperatures. 

\begin{figure*}[ht!]
  \centering 
  \includegraphics[width = \textwidth]{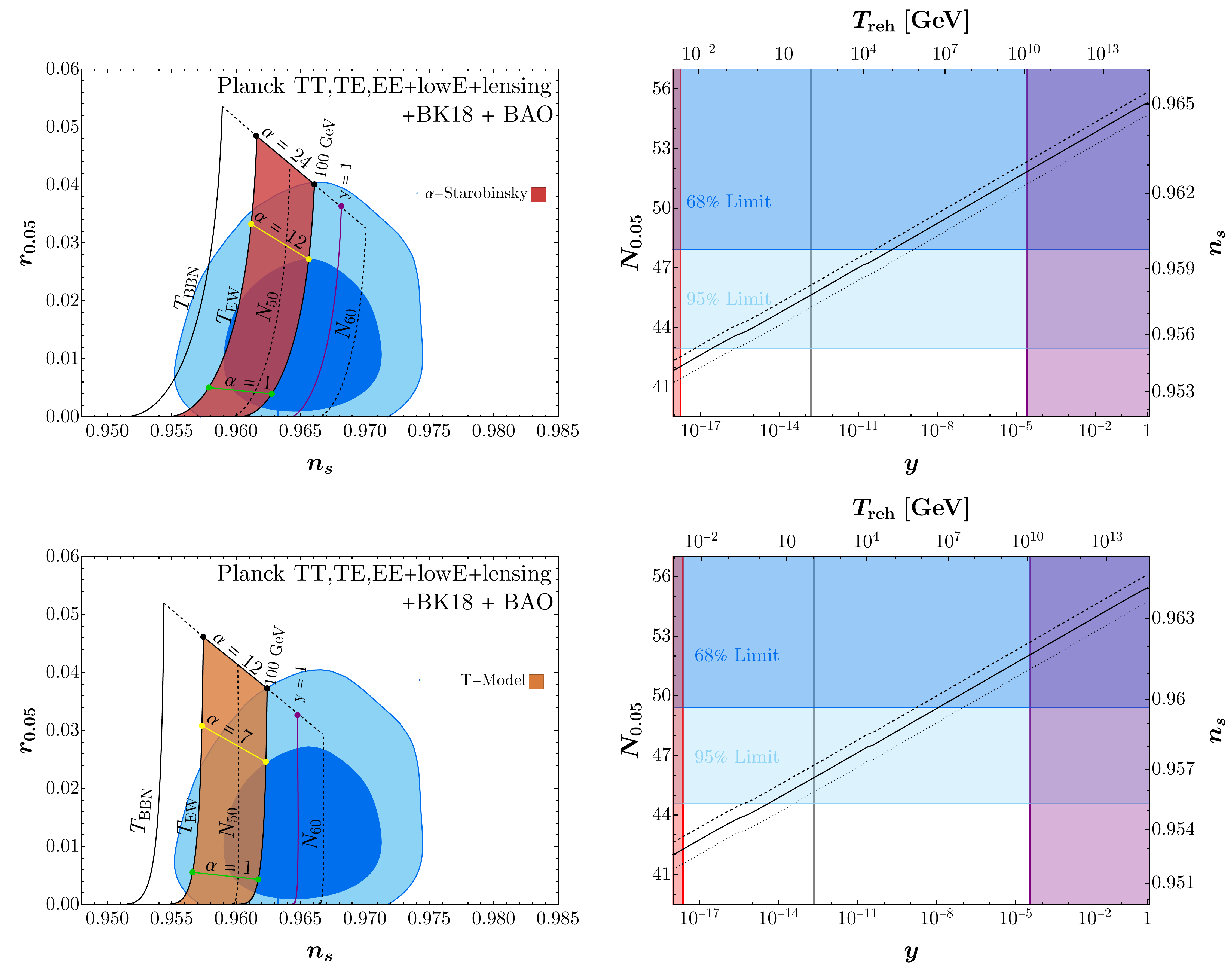}
  \caption{\it Illustrations of the impacts of the BICEP/{\it Keck} and other constraints on $\alpha$-Starobinsky models (upper panels) and T models (lower panels) based on the analysis of \cite{Tristram:2021tvh}. 
  The left panels compare the observational 68\% and 95\% C.L. constraints in the $(n_s, r)$ plane (using pivot scales 0.05 for both $r$ and $n_s$) with the model predictions for different numbers of $e$-folds
  $N_{50, 60}$, showing also the predictions for an inflaton coupling $y = 1$, the  constraints from $T_{\rm reh} \ge T_{\rm BBN}$ and $T_{\rm EW}$,  and the constraints if the LSP mass is $100 \, \rm{GeV}$.
  The right panels display $(y, N)$ planes (using the pivot scale 0.05), showing the relations between
  $y$ and $T_{\rm reh}$ and between $N$and $n_s$, and the values $\alpha = 0.1, 1, 5$ (dashed, solid and dotted black lines). We also include lower limits on $y$ from BBN (red line), 
  $T_{\rm reh} = T_{\rm EW}$ (grey line), and gravitino production (purple line) for $\alpha = 1$, which increase for smaller $\alpha$,
  and 68\% and 95\% C.L. lower limits on $N_{0.05}$ from BICEP/{\it Keck} and other data (blue lines). 
  }
  \label{fig:results2}
\end{figure*}

Fig.~\ref{fig:results2} shows analogous results based on the analysis in \cite{Tristram:2021tvh}. Since this work provides limits on $r$ using 0.05 Mpc$^{-1}$ for the pivot scale,
we have recalculated the theory curves accordingly, although the difference is quite small. What is more striking is the difference in the 68\% and 95\% lower limits to $n_s$. 
These are shifted slightly to smaller values and, as one can see in Fig.~\ref{fig:results2},
a large portion of the red-shaded region (between $T_{\rm EW}$ and the 100 GeV relic density limit) now overlaps the 68\% C.L. observational region (dark blue). In the right panels, 
we see that the weaker lower limits on $n_s$ reduce the lower limits on $N_{0.05}$ and hence allow a smaller inflaton coupling to matter
and a lower reheat temperature. However, the allowed ranges for $\alpha$ are only slightly modified: (0.4, 12) and (0, 24) for the $\alpha$-Starobinsky model at 68\% and 95\% C.L., respectively, and 
(0.5, 7) and (0, 12) for the T model. 

The modified limits analogous to Eqs.~(\ref{eq:limits1}) and (\ref{eq:limits2}) for $\alpha = 1$ are
\begin{align}
\label{eq:limits3}
&\qquad \qquad \text{$\alpha$-Starobinsky}: \nonumber \\
&N_{68 \%}  = 47.9, \quad N_{95 \%} = 42.9,& \nonumber \\
&T_{{\rm{reh}, \,} 68 \%} = 9.8 \times 10^{4} \, {\rm{GeV}}, \quad T_{{\rm{reh}, \,} 95 \%} = 0.031 \, {\rm{GeV}},& \nonumber \\
&y_{68 \%} = 1.9 \times 10^{-10}, \quad y_{95 \%} = 2.8 \times 10^{-17},& \\
&~~& \nonumber \\
\label{eq:limits4}
& \qquad \qquad \text{T Model}: \nonumber \\
&N_{68 \%} = 49.4, \quad N_{95 \%} = 44.6,& \nonumber \\
&T_{{\rm{reh}, \,} 68 \%} = 4.4 \times 10^{6} \, {\rm{GeV}}, \quad T_{{\rm{reh}, \,} 95 \%} = 2.0  \, {\rm{GeV}},& \nonumber \\
&y_{68 \%} = 1.2 \times 10^{-8}, \quad y_{95 \%} = 4.0 \times 10^{-15}.& 
\end{align}

The limits on $N_*$ and $y$ from limits to $T_{\rm reh}$ and the relic density are unaffected by the choice of data analysis and are not repeated. 

\section{Discussion}

As can be seen from the left panels of
Figs.~\ref{fig:results} and \ref{fig:results2}, the primary driver of the
upper limits on $\alpha$ is the new upper limit on $r$,
whereas the constraint on $n_s$ is the primary driver
of the lower limit on the number of $e$-folds. In both
the $\alpha$-Starobinsky and T models there is also
an upper limit on the number of $e$-folds due to requiring
the inflaton decay coupling $y \lesssim {\cal O}(1)$,
namely, $N_* \lesssim 56$ as seen  in the right
panels of the figures, which restricts $n_s$ to
the left halves of the preferred ovals in the left
panels of Figs.~\ref{fig:results} and \ref{fig:results2}. In both cases, couplings near or at this upper limit lead to observables closest to the central value of the confidence contours. This indicates that the updated constraints in $n_s$ favour scenarios for which radiation domination is almost immediately reached after the end of inflation. We note that such a thermal history is always realized regardless of the inflaton-Standard Model couplings if the inflationary potential is quartic near its minimum, as is the case of Higgs inflation \cite{higgsinf}, WIMPflation \cite{wimpflation} or T models 
of the form $V \sim \tanh^{4}(\varphi/\sqrt{6\alpha} M_P)$~\cite{Turner:1983he,GKMO}. For quartic minima, $N_*\simeq 56$, independent of the reheating temperature.

The values of the effective Yukawa coupling $y$ disfavoured by electroweak scale gravitino overproduction, shown in purple in Figs.~\ref{fig:results} and \ref{fig:results2}, 
correspond coincidentally to the domain of non-perturbative particle production (preheating). Indeed, for $y\gtrsim 10^{-5}$, efficient parametric resonance will be present 
during the early stages of reheating, for either fermionic or bosonic inflaton decay products~\cite{Drewes:2015coa,Drewes:2017fmn,Ellis:2019opr,Drewes:2019rxn,GKMOV}. 
However, this effect is not necessarily reflected in the CMB observables. In the case of fermionic preheating, the expansion history during reheating 
(and hence $w_{\rm int}$ and $\rho_{\rm rad}$) is not affected unless $y\sim \mathcal{O}(1)$. The resulting Pauli suppression of particle production 
simply reduces the energy density of radiation relative to the value predicted by (\ref{eq:dyn2}) for a time much shorter than the duration of reheating~\cite{GKMOV}. 
Hence our results for $N_*$ shown in the left panels of Fig.~\ref{fig:results} would be mostly unchanged in this fermionic case. In the case of bosonic preheating, 
the efficiency of non-perturbative particle production depends on the resonance band structure of the coupling. If the backreaction regime is reached, 
transient radiation-dominated stages can occur during reheating, modifying $w_{\rm int}$ and hence our predictions~\cite{Maity:2018qhi,GKMOV}.
However, we do not delve here into this model-dependent issue. Finally, for attractors with quadratic minima, the self-interaction of the inflaton does
not disrupt the matter-like oscillation of the inflaton condensate during reheating~\cite{Lozanov:2017hjm}.

Turning to the future, we note that the experiments CMB-S4~\cite{CMB-S4} and LiteBIRD~\cite{LITEBIRD} 
will target primarily the search for B-modes in the CMB and will impose strong constraints on $r$,
with the potential to reduce substantially the uncertainty in $r$, by a factor ${\cal O}(2)$. Such
a measurement will reduce the uncertainty in $\alpha$ to a similar value,
constraining significantly string models of inflation. 
Unfortunately, the ability of these experiments to constrain $n_s$ is limited.
However, this is an important objective for the future, as $n_s$ is related
directly to the magnitude of the coupling between the inflaton and matter,
whose understanding will be key for connecting the theory of inflation to
laboratory physics.

\subsection*{Acknowledgements}
We thank Marco Drewes, Mathias Pierre, Douglas Scott and Matthieu Tristram for helpful discussions.  The work of J.E. was supported partly by the United Kingdom STFC Grant ST/T000759/1 and partly by the Estonian Research Council via a Mobilitas Pluss grant. J.E., M.A.G.G.~and S.V. acknowledge the hospitality of the Institut Pascal at the Universit{\' e} Paris-Saclay during the 2021 Paris-Saclay Astroparticle Symposium, with the support of the P2IO Laboratory of Excellence program ``Investissements d'avenir" ANR-11-IDEX-0003-01 Paris-Saclay and ANR-10-LABX-0038, the P2I axis of the Graduate School Physics of the Universit\'e
Paris-Saclay, as well as IJCLab, CEA, IPhT, APPEC, and EuCAPT
ANR-11-IDEX-0003-01 Paris-Saclay and ANR-10-LABX-0038. M.A.G.G.~was 
also supported by the IN2P3 master project UCMN. 
The work of D.V.N. was supported partly by the DOE grant DE-FG02-13ER42020 and partly by the Alexander S. Onassis Public Benefit Foundation. The work of K.A.O. was supported in part by DOE grant DE-SC0011842 at the University of Minnesota.

\section*{Appendices}
\appendix
\renewcommand{\thesubsection}{\Alph{subsection}}
\setcounter{equation}{0}
\renewcommand{\theequation}{\thesubsection.\arabic{equation}}
\subsection{Analytical approximations}
\label{appA}

As stated in the main text, the power spectrum and reheating constraints summarized in Fig.~\ref{fig:results} have been obtained numerically. In this Appendix we provide analytical approximations to the relevant inflationary quantities.

The end of inflation corresponds to the end of the epoch of accelerated expansion, i.e., $\ddot{a}=0$ or $\epsilon_H=1$, where $\epsilon_H=-\dot{H}/H^2$ is the first Hubble flow function. In terms of the potential slow-roll parameters (\ref{eq:epseta}), it can be shown that the end of inflation occurs approximately when~\cite{EGNO5}
\beq
\epsilon \;\simeq\; (1+\sqrt{1-\eta/2})^2\,.
\eeq
This expression can be used to obtain the following closed-form
estimates for the value of the inflaton field at the end of inflation for $\alpha$-Starobinsky models,
\begin{align}
\label{eq:phiendE}
\frac{\varphi_{\rm{end}}}{M_P} & \; \simeq \; \sqrt{\frac{3\alpha}{2}} \ln\left[\frac{2(6\alpha + 3\sqrt{3\alpha}-2)}{12\alpha-1} \right] , & 
\end{align}
and for T models,
\begin{align}
\label{eq:phiendT}
\frac{\varphi_{\rm{end}}}{M_P} & \; \simeq \;  \sqrt{\frac{3\alpha}{2}}  \ln\left[ \frac{4-6\sqrt{\alpha(5+4\alpha)}}{1-12\alpha}\right.  & \nonumber\\
& \qquad\qquad\qquad +\left.\sqrt{\frac{75}{5+68\alpha+16\sqrt{\alpha(5+4\alpha)}}}\right] . & 
\end{align}
As expected, for $\alpha = 1$, we recover Eqs.~(\ref{alphastarophiend}) and (\ref{tmodelphiend}). Compared to the exact values, the analytic approximations have errors of 2\% (2\%, 4\%) 
for $\alpha=1$ (0.1, 10) in the case of $\alpha$-Starobinsky models, and of  5\% (3\%, 5\%) 
for $\alpha=1$ (0.1, 10) for T models.

The value of the inflaton field at the moment when the pivot scale crosses the horizon can be 
estimated by integrating Eq.~(\ref{eq:efolds}). In the case of $\alpha$-Starobinsky models, 
\begin{align}\label{eq:phistarstaro} \notag
\frac{\varphi_{*}}{M_P} & \; \simeq \; \sqrt{\frac{3\alpha}{2}}
\left[1 + \frac{3\alpha}{4N_*-3\alpha} \right] \\
& \qquad \times
\ln\left(\frac{4N_*}{3\alpha} + e^{\sqrt{\frac{2}{3}} \frac{\varphi_{\rm{end}}}{M_P}} - \sqrt{\frac{2}{3}} \frac{\varphi_{\rm{end}}}{M_P} \right)  , 
\end{align}
and for T models,
\begin{align}
\label{eq:phistarT}
\frac{\varphi_{*}}{M_P} & \; \simeq \;  \sqrt{\frac{3\alpha}{2}}  \cosh^{-1} \left[\frac{4N_*}{3\alpha} + \cosh\left(\sqrt{\frac{2}{3\alpha}} \frac{\varphi_{\rm{end}}}{M_P} \right) \right]\, .
\end{align}
For $40<N_*<60$ the relative errors are at most
0.3\% (0.3\%, 3\%) for $\alpha=1$ (0.1, 10) in the $\alpha$-Starobinsky case, and 
0.5\% (0.4\%, 0.7\%) for $\alpha=1$ (0.1, 10) in the case of T model inflation.

The logarithm of the so-called reheating parameter~\cite{Martin:2010kz},
\begin{align}\label{eq:lnrrad}
\ln R_{\rm rad} \;&\equiv\; \ln\left[\frac{a_{\rm end}}{a_{\rm rad}}\left(\frac{\rho_{\rm end}}{\rho_{\rm rad}}\right)^{1/4}\right]\\  
&=\; \frac{1-3w_{\rm int}}{12(1+w_{\rm int})}\ln\left(\frac{\rho_{\rm rad}}{\rho_{\rm end}}\right)\,,
\end{align}
may be estimated by noting that the energy density of the relativistic inflaton decay products,
assuming a constant decay rate $\Gamma_{\varphi}$, can be written as~\cite{EGNO5}:
\beq\label{eq:rhorad}
\rho_{\rm rad} \;=\; \rho_{\rm end}\left(\frac{a_{\rm end}}{a_{\rm rad}}\right)^4\int_{0}^{v_{\rm rad}}\left(\frac{a(u)}{a_{\rm end}}\right)e^{-u}du\,,
\eeq 
where $v\equiv \Gamma_{\varphi}(t-t_{\rm end})$. Approximating the equation-of-state parameter as $w \simeq 0$ during reheating, we can further write
\begin{align} \label{eq:aforreh}
\frac{a(t)}{a_{\rm end}} \;&\simeq\;  \left(\sqrt{\frac{3}{4}\rho_{\rm end}}\frac{t-t_{\rm end}}{M_P}\right)^{\frac{2}{3}} \;=\; \left(\frac{3H_{\rm end}v}{2\Gamma_{\varphi}}\right)^{\frac{2}{3}}\,.
\end{align}
Substitution of (\ref{eq:aforreh}) into (\ref{eq:rhorad}) and subsequently into (\ref{eq:lnrrad}) results in the following simple approximation for the reheating parameter,
\beq\label{eq:logRradf}
\ln R_{\rm rad} \;\simeq\; \frac{1}{6}\ln\left(\frac{\Gamma_{\varphi}}{H_{\rm end}}\right)\,.
\eeq
This result allows us to write simple analytical expressions for the number of $e$-folds
after horizon crossing as functions of the effective Yukawa coupling responsible for reheating. As an example for $\alpha=1$, substitution of (\ref{eq:phiendE}), (\ref{eq:phistarstaro}) and (\ref{eq:logRradf}) into (\ref{eq:nstarreh}) gives
\begin{align} 
\label{eq:nEmodapprox}
N_* \;&\simeq\; 57.68 -\frac{1}{2}\ln N_* + \frac{1}{3}\ln y - \frac{1}{12}\ln g_{\rm reh} \,,
\end{align}
for $\alpha$-Starobinsky models at the pivot scale $k_*=0.05\,{\rm Mpc}^{-1}$, and for T models
\begin{align} 
\label{eq:nTmodapprox}
N_* \;&\simeq\; 57.82 -\frac{1}{2}\ln N_* + \frac{1}{3}\ln y - \frac{1}{12}\ln g_{\rm reh} \,.
\end{align}
In the range of values shown in the left panels of Fig.~\ref{fig:results}, the maximum differences of these approximations from the full numerical results are 0.2\% (0.1\%) for the $\alpha$-Starobinsky models (T models).

For other analyses of reheating in attractor models, see \cite{Drewes:2017fmn,German}.

\subsection{Computing the CMB observables}
\label{appB}
\setcounter{equation}{0}

In order to compute accurately the inflationary observables, in particular the scalar tilt $n_s$, 
we have integrated the linear equations for the curvature fluctuation numerically. 
To calculate the gauge-invariant Mukhanov-Sasaki variable $Q$,\footnote{In the Newtonian gauge, $Q = \delta\varphi + \frac{\dot{\varphi}}{H}\Psi$, where $\delta\varphi$ and $\Psi$ denote the field and the metric perturbations, respectively.} we integrate the equation of motion \cite{Lalak:2007vi,egno3},
\beq\label{eq:MSeq}
\ddot{Q} + 3H\dot{Q} + \left[ \frac{k^2}{a^2} + 3\dot{\varphi}^2 - \frac{\dot{\varphi}^4}{2H^2} + 2\frac{\dot{\varphi}V_{\varphi}}{H} + V_{\varphi\varphi} \right] Q \;=\;0\,,
\eeq
with the Bunch-Davies initial condition $Q_{k\gg aH} = e^{-ik\tau}/a\sqrt{2k}$, 
where $d\tau = dt/a$ is the conformal time. The corresponding metric fluctuation and its power spectrum are in turn given by
\begin{align}
\mathcal{R} \;&=\; \frac{H}{|\dot{\varphi}|}Q\,,\\
\langle \mathcal{R}(k) \mathcal{R}^*(k') \rangle \;&=\; \frac{2\pi^2}{k^3}\mathcal{P}_{\mathcal{R}} \delta (k-k')\,.
\end{align}
The scalar tilt is then computed using its definition,
\beq\label{eq:nsdef}
n_s \;=\; 1 + \frac{d \ln\mathcal{P}_{\mathcal{R}}}{d\ln k}\, ,
\eeq
and the tensor-to-scalar-ratio is
\begin{equation}
    \label{eq:rdef}
    r \; = \; \frac{\mathcal{P}_{\mathcal{T}}}{\mathcal{P}_{\mathcal{R}}} \, ,
\end{equation}
where in the case of the tensor spectrum we take the horizon-crossing value $\mathcal{P}_{\mathcal{T}}= 2H^2/\pi^2$.

\begin{figure}[ht!]
  \centering 
  \includegraphics[width = \columnwidth]{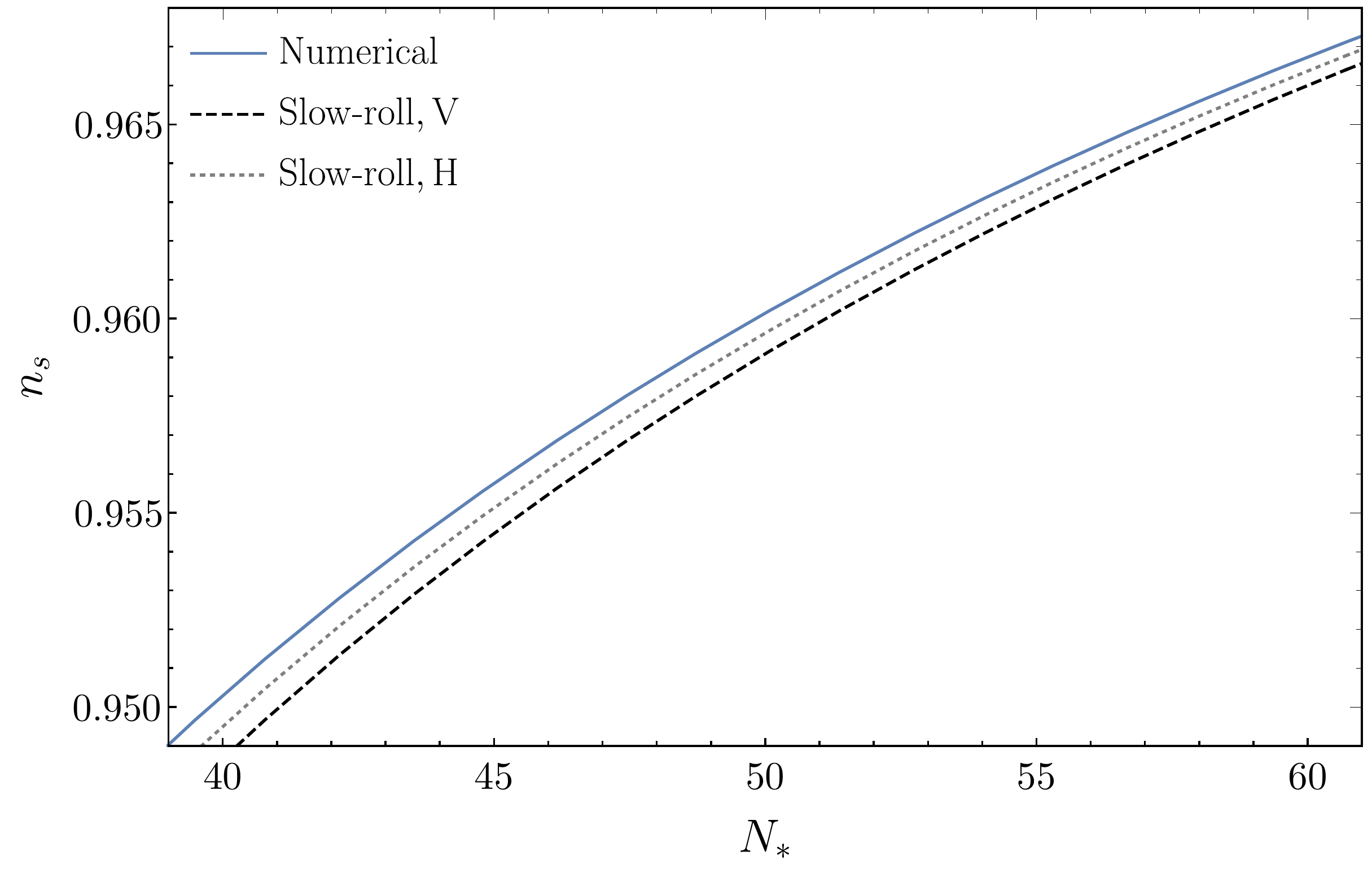}
  \caption{\it The scalar tilt $n_s$ as a function of the number of $e$-folds after horizon crossing,
  $N_*$, for the $\alpha$-Starobinsky model with $\alpha=1$. The continuous blue line is the 
  numerical solution of Eqs.~(\ref{eq:MSeq})-(\ref{eq:nsdef}). The dotted grey line is the slow-roll approximation (\ref{eq:spectrtilt}) with the Hubble parameters $\epsilon_H,\eta_H$ defined in (\ref{eq:epsetaH}). The dashed black line is
  the slow-roll approximation (\ref{eq:spectrtilt}) calculated using the potential parameters 
  $\epsilon,\eta$ defined in (\ref{eq:epseta}). }
  \label{fig:nsN}
\end{figure}

Comparing the numerical results obtained by the procedure above with the slow-roll approximations (\ref{eq:spectrtilt}) and (\ref{eq:sclrtotens}) we find a discrepancy $\gtrsim 1$ $e$-fold for $N_*=N_*(n_s)$, 
see the dashed line in Fig.~\ref{fig:nsN}. This difference can be reduced if, instead of the potential 
slow-roll parameters (\ref{eq:epseta}) one uses the Hubble slow-roll parameters, 
\beq\label{eq:epsetaH}
\epsilon_H \;=\; -\frac{\dot{H}}{H}\,, \qquad \eta_{H} \;=\; 2\epsilon_H - \frac{\dot{\epsilon_H}}{2\epsilon_H H}\,,
\eeq
\\
see the dotted line in Fig.~\ref{fig:nsN}.

This difference remains even when the higher-order slow-roll corrections are included.
Ultimately, it is due to the fact that curvature modes do not immediately freeze 
upon leaving the horizon, which corresponds to the condition $k=aH$. Hence there is 
always a shift between the approximate horizon-crossing value, used in our 
semi-analytical estimates, and the final ``freeze-out'' values used in our full numerical results, 
in particular in Fig.~\ref{fig:results}.

\end{document}